# A Non-Abelian Route to $Z_2$ Non-Hermitian Skin Effects

Huiyan Tang[1†], Yaxuan Zhang[2†], Ziteng Wang[1†], Liqin Tang[1], Daohong Song[1], Jingjun Xu[1], Weixuan Zhang[2*], Hrvoje Buljan[1,3], Xiangdong Zhang[2*], and Zhigang Chen[1*]

[1]*TEDA Applied Physics Institute and School of Physics, Nankai University, Tianjin 300457, China*
[2]*School of Physics, Beijing Institute of Technology, Beijing 100081, China*
[3]*Department of Physics, Faculty of Science, University of Zagreb, Bijenička Cesta 32, 10000 Zagreb, Croatia*
[†]*These authors contributed equally to this work*

*e-mail : zhangwx@bit.edu.cn, zhangxd@bit.edu.cn, zgchen@nankai.edu.cn

## Abstract

The non-Hermitian skin effect (NHSE), characterized by extensive boundary accumulation of eigenstates under open boundary conditions, has emerged as a central phenomenon in non-Hermitian physics. Conventionally, the NHSE arises from either non-reciprocal couplings or onsite gain and loss combined with synthetic gauge fields. Existing studies, however, have been largely confined to frameworks with Abelian-coupling, leaving the role of non-Abelian couplings essentially unexplored. Here, we demonstrate that non-Abelian-couplings can generate the NHSE, giving rise to a time-reversal-symmetry-protected $Z_2$ skin effect with pseudospin-dependent boundary localization and dynamical pseudospin separation. Experimentally, we implement a representative four-level model using a programmable topolectrical circuit and directly observe both the predicted NHSE and the boundary-induced pseudospin-inversion reflection. Our work establishes a fundamental link between non-Abelian coupling and non-Hermitian topology, opening new avenues for realizing non-reciprocity-free topological materials and devices.

**Keywords:** non-Abelian-coupling, $Z_2$ topology, non-Hermitian skin effect, pseudospin-inversion, topolectrical circuit

Recent years have witnessed the emergence of non-Hermitian physics as a central paradigm in condensed matter physics and optics, driven by the recognition that gain, loss, and dissipation can fundamentally reshape wave dynamics and spectral topology [1-4]. Beyond the conventional Hermitian paradigm, non-Hermitian Hamiltonians generally exhibit complex energy spectra and non-orthogonal eigenstates, giving rise to a variety of phenomena without Hermitian counterparts, including exceptional points (EPs) [2,4], non-Hermitian topological phases [5-7], and unconventional wavefunction dynamics [8,9]. A particularly striking manifestation is the non-Hermitian skin effect (NHSE) [10-26], whereby an extensive number of eigenstates accumulate at system boundaries under open boundary conditions (OBC), in sharp contrast to the conventional bulk-boundary correspondence. To date, the NHSE has been predominantly explored in systems governed by Abelian forms of non-Hermiticity, arising either from nonreciprocal couplings [10,22,23] or from synthetic gauge fields combined with gain and loss [20,21]. Recently, a time-reversal-symmetry (TRS)-protected $Z_2$ NHSE [12,20] has been demonstrated [25,26]. Despite these advances, the role of non-Abelian couplings in non-Hermitian systems, particularly their influence on boundary sensitivity and symmetry constraints, remains poorly understood.

Non-Abelian theory, which underpins the description of non-commutative relations in quantum and wave systems, has long occupied a central position in modern physics [14,27-42]. In lattice and synthetic-matter platforms, non-Abelian gauge fields act on internal degrees of freedom such as spin or polarization, thereby enriching band structure and transport beyond Abelian descriptions and giving rise to phenomena including the non-Abelian Aharonov-Bohm effect [34,36] and quantum Hall physics [31]. This framework provides a powerful route to engineering geometric phases, matrix-valued holonomies, and symmetry-protected dynamics that have no Abelian analogues. While recent experiments in metamaterials [37], transmission line networks [38], acoustic semimetals [39], topolectrical circuits [33], optical fiber networks [34,35], and waveguide lattices [41,42] have successfully demonstrated a variety of non-Abelian phenomena, these studies have largely remained within Hermitian settings. By contrast, how non-Abelian couplings modify the emergence, topology, and dynamics of the NHSE, and whether they introduce qualitatively new physical principles, is still an open and largely unexplored question.

Here, we introduce and experimentally realize a $Z_2$ NHSE enabled by non-Abelian couplings without requiring nonreciprocal couplings or gain-loss-induced gauge fields, revealing a distinct route to the NHSE. We theoretically propose a generalized zigzag-chain model representing a four-level spinful system with non-Abelian couplings. As illustrated in Fig. 1, when the couplings take a non-Abelian form, the system exhibits pseudospin-dependent boundary localization; by contrast, in the Abelian limit all modes remain extended. We further confirm the TRS protection of the effect by intentionally breaking TRS locally at a boundary, which induces pseudospin inversion and boundary reflection. Experimentally, using programmable topolectrical circuits that implement non-commuting couplings, we directly observe the predicted pseudospin dynamics. Together, these results establish a pseudospin-resolved $Z_2$ NHSE and provide a platform for exploring previously inaccessible boundary and dynamical phenomena in non-Hermitian topological systems.

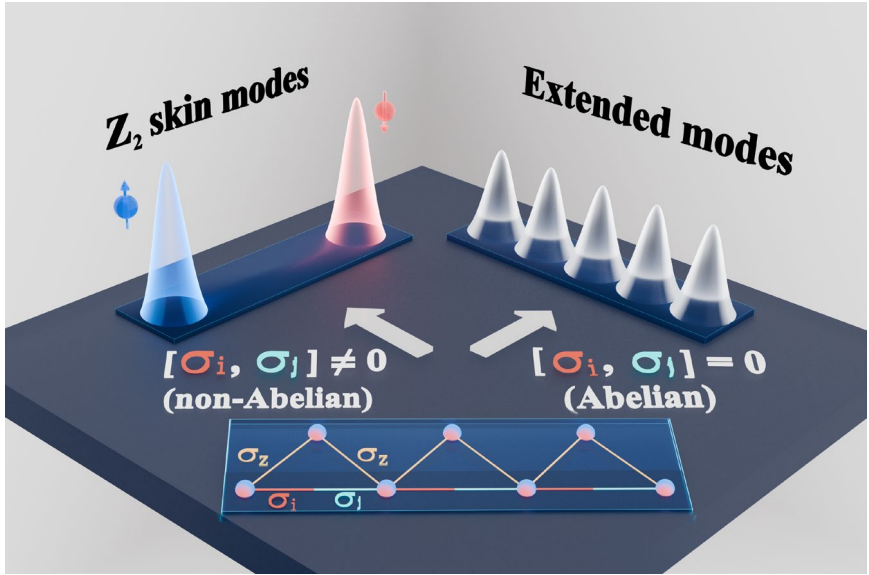

FIG 1. Schematic illustration of $Z_2$ non-Hermitian skin effect under non-Abelian coupling. In our non-Abelian zigzag-chain model, each lattice site hosts two pseudospin degrees of freedom, represented by the red and blue colors. Inter-site hopping is governed by matrix-valued coupling terms $\sigma_z$, $\sigma_i$ and $\sigma_j$ (with $i,j \in (x,y,z)$). When the couplings are non-Abelian ($[\sigma_i, \sigma_j] \neq 0$), the system can exhibit a $Z_2$ NHSE characterized by pseudospin-resolved boundary localization. In contrast, for Abelian-coupling case ($[\sigma_i, \sigma_j] = 0$), the eigenstates remain extended across the lattice. Bloch-sphere illustrations of non-Abelian path dependence are provided in the Supplementary Material S1.

*Non-Abelian-coupling induced $Z_2$ NHSE:* As a representative realization, we show that non-Abelian couplings can generate a symmetry-protected skin effect even in lattices with fully reciprocal hopping. We investigate the role of non-Abelian couplings in non-Hermitian topology by considering the following four-level lattice model, which is constructed on a one-dimensional zigzag lattice with alternating onsite gain and loss (see Fig. 2(a)). Each sublattice (A and B) contains two spinors, represented by the basis states $|1\rangle$ and $|2\rangle$ from the following creation operators:

$$|1\rangle_A = a_1^\dagger |0\rangle, \qquad |2\rangle_A = a_2^\dagger |0\rangle, \qquad |1\rangle_B = b_1^\dagger |0\rangle, \qquad |2\rangle_B = b_2^\dagger |0\rangle. \qquad (1)$$

In this representation, the couplings between sublattices A and B are represented by $t_1\sigma_z$, $t_2\sigma_z$ and $t_3\sigma_x\sigma_z$, where $t_1, t_2, t_3$ are coupling strengths and $\sigma_x, \sigma_z$ are the Pauli matrices. The Hamiltonian is given by:

$$H = \sum_{n=1}^{N}\left(t_1 a_n^\dagger \sigma_z b_n - t_2 b_n^\dagger \sigma_z a_{n+1} + t_3 a_n^\dagger \sigma_x \sigma_z a_{n+1} + \text{H.c.}\right) + i\gamma_1 a_n^\dagger a_n - i\gamma b_n^\dagger b_n. \quad (2)$$

Here $a_n = [a_{n,1}, a_{n,2}]^T$ denotes the annihilation operators for the basis states on sublattice A in the $n$th unit cell (with the analogous definition for $b_n$ on sublattice B), $N$ is the total number of unit cells, and $\gamma_1$, $\gamma$ represent the onsite gain and loss strengths, respectively. All hopping processes in this model are reciprocal; the emergence of the skin effect does not rely on non-reciprocal couplings, and the non-Abelian character arises from the non-commutativity between different coupling matrices.

The non-Abelian couplings imply that a particle may take two (or more) pathways in which the final spin states are unequal [35]. In our lattice model (Fig. 2(a)), starting from an initial state $\psi_0$ at site A, a clockwise coupling pathway yields $\psi^{\text{CW}} = (t_3\sigma_x)(t_3\sigma_z)(-t_2\sigma_z)(t_1\sigma_z)\psi_0$, whereas an counterclockwise pathway gives $\psi^{\text{CCW}} = (t_1\sigma_z)(-t_2\sigma_z)(t_3\sigma_z)(t_3\sigma_x)\psi_0$. These two final spin states are different because $[\sigma_z, \sigma_x] \neq 0$, demonstrating the non-Abelian nature of couplings. Bloch-sphere illustrations are provided in the Supplementary Material S1 [43].

To characterize the skin dynamics, we define the pseudospin basis states as linear combinations of the sublattice spinors (see Supplementary Material S2 and S3 [43]):

$$\begin{aligned} |\uparrow\rangle_A &= \tfrac{1}{2}(|1\rangle_A + |2\rangle_A), \\ |\uparrow\rangle_B &= \tfrac{1}{2}(|1\rangle_B - |2\rangle_B), \\ |\downarrow\rangle_A &= \tfrac{1}{2}(|1\rangle_A - |2\rangle_A), \\ |\downarrow\rangle_B &= \tfrac{1}{2}(|1\rangle_B + |2\rangle_B). \end{aligned} \quad (3)$$

with corresponding projectors defined by:

$$P_\uparrow = |\uparrow\rangle_A\langle\uparrow|_A + |\uparrow\rangle_B\langle\uparrow|_B, \qquad P_\downarrow = |\downarrow\rangle_A\langle\downarrow|_A + |\downarrow\rangle_B\langle\downarrow|_B. \quad (4)$$

By calculating the expectation values of these operators for the eigenstates $|\psi_n\rangle$ of the Hamiltonian, we define the pseudospin projection weights as:

$$n_\uparrow = \langle\psi_n|P_\uparrow|\psi_n\rangle, \qquad n_\downarrow = \langle\psi_n|P_\downarrow|\psi_n\rangle. \quad (5)$$

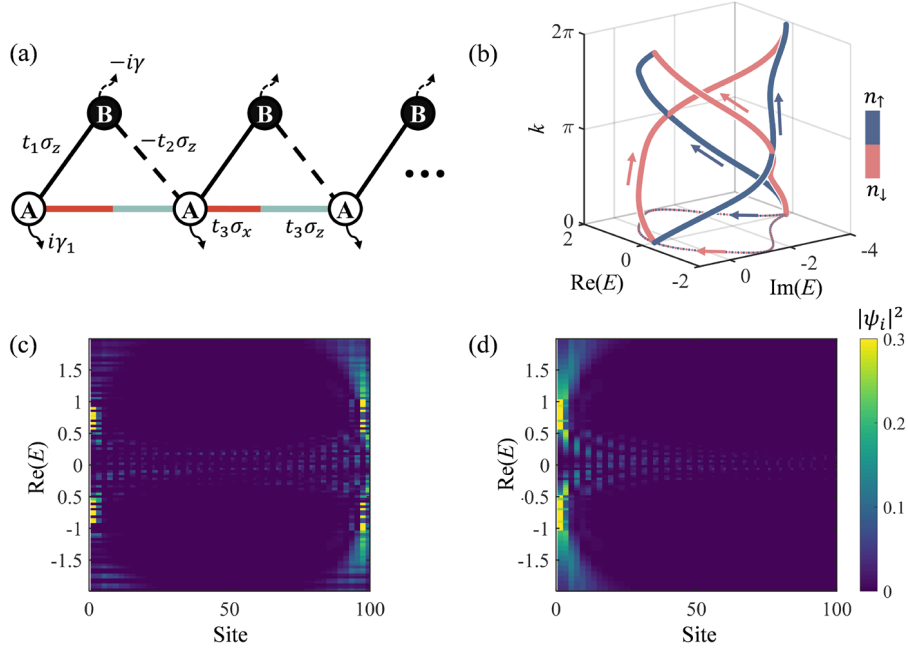

FIG 2. $Z_2$ NHSE in a non-Abelian zigzag lattice. (a) Schematic of the four-level zigzag model with reciprocal non-Abelian couplings. (b) Complex eigenvalues under PBC. Red and blue bands correspond to eigenstates with $n_\downarrow = 1$ and $n_\uparrow = 1$, respectively, as defined in Eq. (5). Red and blue arrows denote clockwise and counterclockwise winding directions of the complex energy spectrum. (c)-(d) Profiles of the eigenmodes of the model under OBC, with boundaries: (c) both preserving TRS and (d) the right boundary breaking TRS via a perturbation $\Delta = \text{diag}(0,0.5)$. The color intensity represents the probability density normalized such that $\sum_i |\psi_i|^2 = 1$. Parameters are $t_1 = 2$, $t_2 = 1$, $t_3 = 0.5$, $\gamma = 4$ and $\gamma_1 = 0.8$.

The momentum-space band structure is plotted in Fig. 2(b), where each complex eigenvalue forms a Kramers pair with opposite winding numbers. Specifically, our numerical analysis shows that the clockwise-winding energy bands (red) feature $(n_\uparrow, n_\downarrow) = (0,1)$, whereas the counterclockwise-winding bands (blue) have $(n_\uparrow, n_\downarrow) = (1,0)$. This pseudospin polarization implies that pseudospin-up modes accumulate at the left boundary, while pseudospin-down modes localize at the right boundary under OBC. This behavior is enforced by TRS, which maps a left-localized state $|\beta, E, \uparrow\rangle$ ($\beta < 1$) to its right-localized Kramers partner $|1/\beta, E, \downarrow\rangle$ ($1/\beta > 1$). Here $\beta = e^{ik}$ ($k \in \mathbb{C}$) and $E$ is the eigenvalue (see Supplementary Material S2 [43]). The resulting bidirectional skin effect with opposite pseudospins [Fig. 2(c)] constitutes the defining signature of the $Z_2$ NHSE.

To demonstrate the symmetry protection, we introduce a TRS-breaking perturbation at the right boundary and modify the coupling to $t_R = t_1 \sigma_z - \Delta$. As illustrated in Fig. 2(d), this symmetry breaking can lead to a sharp transition from bidirectional to unidirectional skin localization. Conversely, numerical results presented in Supplementary Material S4 [43] show that both TRS and non-Abelian couplings are essential: the $Z_2$ skin modes are robust against perturbations only when these two conditions are simultaneously satisfied.

*Pseudospin-Dependent Dynamics*: The wave dynamics exhibit a sensitive dependence on pseudospin, providing a direct visualization of the $Z_2$ NHSE. We consider a Gaussian wave packet centered at $k = 0$ as the initial excitation. Because this initial state is a coherent superposition of pseudospins (indicated by the purple profile at $t = 0$), the subsequent evolution exhibits a spontaneous spatial separation of its pseudospin components. When TRS is preserved at the boundaries, the wave packet splits into two counter-propagating branches, as shown in Fig. 3(a). The pseudospin-up component (blue) propagates leftward, whereas the pseudospin-down component (red) propagates rightward. This spontaneous pseudospin separation constitutes a characteristic signature of the $Z_2$ NHSE.

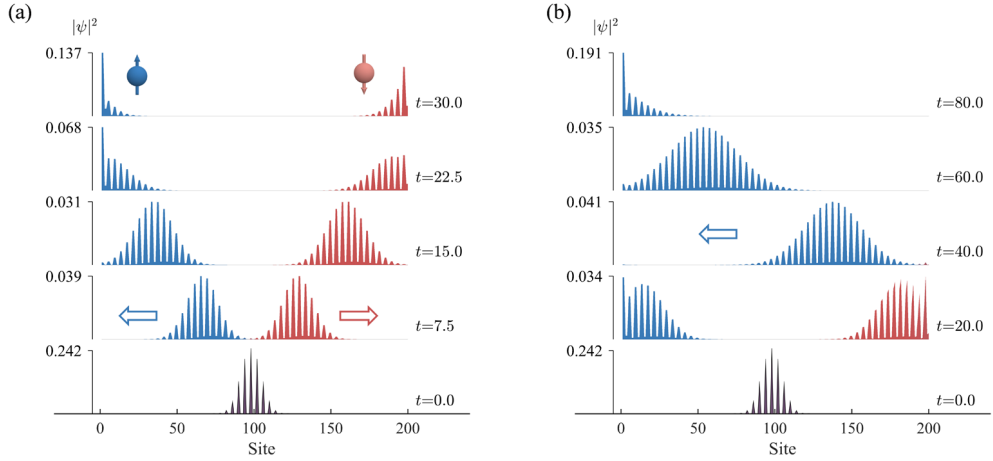

FIG 3. Boundary-controlled wave-packet dynamics and pseudospin inversion. (a, b) Time evolution of a Gaussian wave packet centered at momentum $k = 0$. While the bulk Hamiltonian parameters are identical in both cases, the boundary conditions differ: (a) TRS is preserved at both edges. (b) TRS is broken at the right edge, inducing complete reflection accompanied by pseudospin inversion. The normalized density profiles $|\psi|^2$ are shown at different time steps, with the vertical axes indicating the peak intensities. In the plots, the narrow arrows indicate the propagation direction of the wave packet, while the sphere-headed arrows represent the pseudospin state ($|\uparrow\rangle$ or $|\downarrow\rangle$).

Intuitively, the pseudospin-up states correspond to in-phase modes on sublattice A and out-of-phase modes on sublattice B, whereas the pseudospin-down states correspond to out-of-phase modes on sublattice A but in-phase modes on sublattice B [Eq. (3)]. These dynamical behaviors are fully consistent with the theoretical predictions for a $Z_2$ NHSE. Although the system is intrinsically bidirectional, effectively unidirectional skin transport can be achieved by engineering the initial excitation. By tailoring the initial state to match a specific pseudospin sector, the counter-propagating branch can be selectively suppressed (see Supplementary Material S5 [43]).

The pseudospin-dependent bidirectional dynamics are protected by TRS. Breaking TRS at a boundary introduces both pseudospin inversion and mode reflection. As shown above, breaking TRS at the system's right boundary converts the initially bidirectional skin states into purely left-localized skin states [Figs. 2(c) and 2(d)]. In this scenario, the initial wave packet continues to split into left- and right-moving parts (the bulk dynamics remains unchanged), whereas the rightward-propagating wave packet is fully reflected upon encountering the TRS-broken boundary and eventually localizes at the system's left boundary [see Fig. 3(b)]. This reflection process is accompanied by pseudospin inversion, whereby the rightward-propagating pseudospin-down states are converted into the pseudospin-up states. This combined mechanism of mode reflection and pseudospin inversion highlights the interplay between TRS-breaking and the $Z_2$ NHSE, revealing a new dynamical control in non-Hermitian systems.

*Experimental Observation of $Z_2$ NHSE Dynamics:* Motivated by recent experimental implementations of topological lattice models using circuit networks [32,33,44-54], we construct a programmable topolectrical circuit designed to directly observe the $Z_2$ NHSE dynamics. As depicted in Fig. 4(a), circuit lattices provide a powerful platform in which the Laplacian admittance matrix is mathematically isomorphic to the tight-binding Hamiltonian $H$ (see Supplementary Material S3 for details [43]). Our design maps the four-component spinor basis of the unit cell onto four independent circuit nodes, with node voltages governed by Kirchhoff's laws. For example, a representative unit cell (gray shaded region) comprises four nodes with voltages labeled as $V_n^{a,1}$, $V_n^{a,2}$, $V_n^{b,1}$, and $V_n^{b,2}$, corresponding to the internal degrees of freedom of sublattices A and B in the $n$th unit cell. This architecture establishes a direct experimental correspondence between the non-Abelian coupling matrices of the model and measurable voltage dynamics in the circuit.

The realization of non-Abelian-coupled non-Hermiticity relies on the precise engineering of active circuit elements. The circuit consists of four functional elements: onsite gain at sublattice A, onsite loss at sublattice B, and reciprocal intra- and inter-cell hoppings. The onsite gain and loss are implemented via grounded current-inversion negative-impedance converters (INICs) and resistors to ground, respectively. As detailed in the inset of Fig. 4(a), the INICs provide an equivalent negative resistance $-R_G$ (gain), while standard resistors $R_L$ provide dissipation (loss). Crucially, the matrix-valued hopping terms are realized through resistive networks bridging the circuit nodes. Intra- and inter-cell couplings are realized by $\pm R_i$ ($i = 1,2,3$). To implement the negative couplings, we utilize INICs to synthesize bidirectional negative resistances. This configuration enables precise control of the hopping signs ($\pm t$) while strictly preserving reciprocity, in contrast to schemes relying on explicitly non-reciprocal active components.

The circuit dynamics are governed by Kirchhoff's laws, which map to the target Hamiltonian [Eq. (2)]. The quantitative relationship between the model parameters ($t_i$, $\gamma_1$, $\gamma$) and the circuit component values ($R, C$) are established as:

$$t_i = \frac{\delta}{CR_i} (i = 1,2,3), \qquad \gamma_1 (\text{or } \gamma) = \frac{\delta}{CR_{G/L}}, \tag{6}$$

where $\delta$ is a global scaling factor, and $C$ represents the nodal capacitance. Detailed circuit schematics, effective grounding corrections, and complete parameter derivations are provided in Supplementary Material S3 and S7 [43].

We construct two topolectrical circuits of identical length ($N = 13$) representing the lattice under preserved and broken TRS, respectively. The experimental setup is shown in Fig. 4(b), where the red dashed box highlights the right boundary of the system, and the inset displays a magnified view of the boundary coupling configuration. We first examine the TRS-preserving case by exciting the central node and measuring the time-resolved voltage response (see Supplementary Material S6 for details [43]). The evolution profile [Fig. 4(c)] demonstrates the splitting of the initial excitation into counter-propagating skin modes. Figures 4(e) and 4(f) confirm that the $|\uparrow\rangle$ components (in-phase modes on sublattice A and out-of-phase modes on sublattice B) undergo leftward skin-mode propagation, whereas the $|\downarrow\rangle$ components (out-of-phase modes on sublattice A and in-phase modes on sublattice B) propagate rightward, in full agreement with the theoretical predictions. These measurements provide a direct experimental visualization of the pseudospin-resolved nature of the $Z_2$ NHSE.

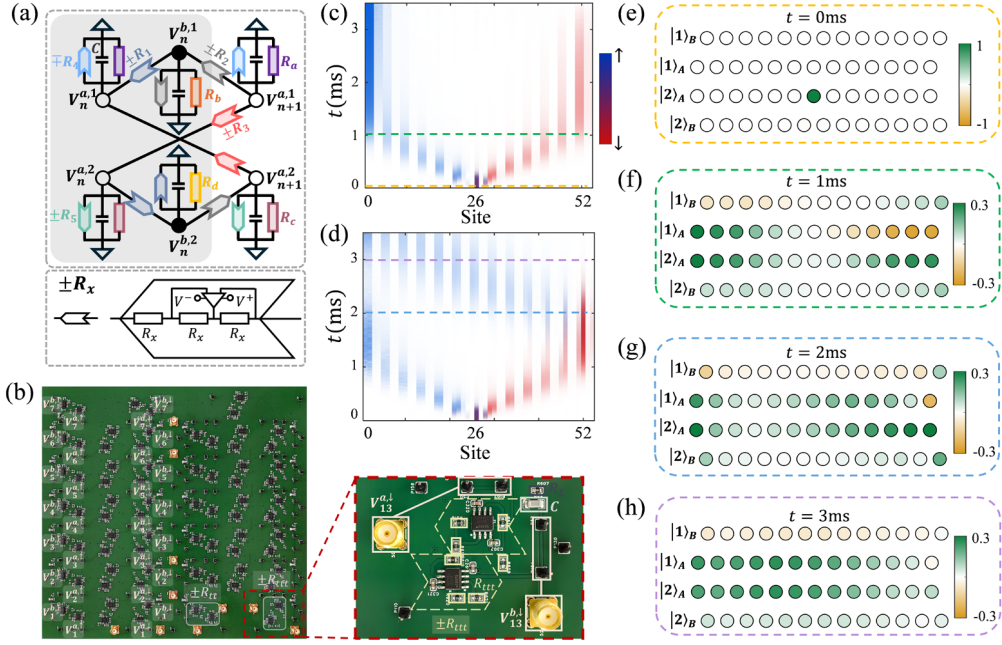

FIG 4. Experimental observation of $Z_2$ NHSE dynamics in topological circuits. (a) Local schematic of the topolectrical circuit realizing the $Z_2$ NHSE. Colors denote distinct coupling modules and resistor values. The gray shaded region denotes a single unit cell. The lower inset illustrates the INIC-based coupling element. (b) Photograph of the fabricated circuit board. The dashed box highlights the boundary region. The right panel shows a magnified view of the boundary coupling configuration. (c, d) Dynamical evolution following a single-sublattice excitation. The two circuit configurations share most component values: $C = 100$ nF, $R_1 = 1$ kΩ, $R_2 = 2$ kΩ, $R_3 = 4.02$ kΩ, and $R_L = 499$ Ω, with reciprocal elements $\pm R_1 = \pm 1$ kΩ, $\pm R_2 = \pm 2$ kΩ, and $\pm R_3 = \pm R_G = \pm 4.02$ kΩ. The only difference lies at the boundaries: (c) TRS is preserved at both ends, with $\pm R_{R,1} = \pm R_{R,2} = \pm R_1$. (d) TRS is broken at the right edge, leading to reflection and pseudospin inversion, implemented via $\pm R_{R,1} = \pm 1$ kΩ, $\pm R_{R,2} = \pm 1.33$ kΩ. Colored horizontal dashed lines indicate the time slices shown in (e)-(h). (e-h) Experimental pseudospin-resolved voltage distributions at $t = 0$ms, 1ms, 2ms, and 3ms, showing the sequence of initial excitation, propagation, edge localization, TRS-breaking-induced reflection, and pseudospin inversion.

To probe the interplay between non-Hermitian topology and symmetry breaking, we mechanically reconfigure the right boundary to break TRS while keeping the bulk parameters unchanged. Although the initial bulk splitting remains unaltered, the boundary dynamics are profoundly modified. As shown in Fig. 4(d), the rightward-propagating wave packet undergoes complete reflection upon reaching the TRS-broken edge and eventually localizes at the left boundary. The microscopic snapshots in Figs. 4(g) and 4(h) reveal a pseudospin-inversion scattering process, whereby the incident pseudospin-down modes are converted into reflected pseudospin-up modes. By resolving the time-resolved voltage dynamics of individual nodes, we directly visualize the interplay between TRS breaking and non-Hermitian topology, providing clear evidence that boundary symmetry breaking can control the topology of skin modes.

The role of non-Abelian couplings in non-Hermitian topology remains largely unexplored, as most non-Hermitian topological models have been formulated within Abelian-coupling frameworks. Here we show that a $Z_2$ NHSE can emerge in a lattice with non-Abelian couplings, as demonstrated through our theoretical model and experimental topolectrical realization. Importantly, the original proposal of the $Z_2$ NHSE [12], as well as its subsequent experimental realizations [25,26], did not consider non-Abelian coupling. While the $Z_2$ NHSE is now recognized as a non-Hermitian topological phase, we do not claim that non-Abelian coupling is either a necessary or sufficient condition for its existence. Instead, our work identifies non-Abelian coupling as a new and versatile degree of freedom for engineering non-Hermitian topological phenomena, offering new design principles for constructing lattice systems with enriched boundary and dynamical properties.

In summary, we have theoretically established and experimentally demonstrated a $Z_2$ NHSE in a non-Abelian-coupled lattice. We introduced a four-level spinful lattice model in which non-Abelian couplings give rise to pseudospin-resolved boundary accumulation of skin modes. We further uncovered a distinctive dynamical signature: local breaking of time-reversal symmetry (TRS) at a boundary induces a pseudospin-inversion reflection process that converts counter-propagating skin modes into unidirectional localization. This mechanism was directly visualized using a programmable topolectrical circuit, enabling real-time access to pseudospin-resolved non-Hermitian dynamics. By extending non-Hermitian topology to lattices with non-Abelian couplings, our results establish a general framework for symmetry-controlled skin effects in reciprocal systems, opening new avenues for robust non-reciprocity-free control of wave transport in both classical and quantum platforms. More broadly, these results suggest that non-Abelian internal degrees of freedom can fundamentally reshape non-Hermitian boundary physics beyond conventional reciprocity-breaking paradigms.

**Acknowledgments**

This research was supported by the National Key R&D Program of China (No. 2022YFA1404800); the National Natural Science Foundation of China (No. W2541003, 12134006, 124B2078, 12374309); the 111 Project (No. B23045) in China. H.B. acknowledges support from the project "Implementation of cutting-edge research and its application as part of the Scientific Center of Excellence for Quantum and Complex Systems, and Representations of Lie Algebras", Grant No. PK.1.1.10.0004, co-financed by the European Union through the European Regional Development Fund—Competitiveness and Cohesion Programme 2021-2027.

**Date Availability**

All source data that support the plots within this paper and other findings of this study are available from the corresponding authors upon reasonable request.

and TRS-induced mapping, details on the topolectrical circuit realization and Hamiltonian isomorphism, and extensive data on the robustness and experimental measurements.